\documentclass[useAMS]{mn2e}
\input{psfig}
\usepackage{times}

\def\Real{{\rm I\mathchoice{\kern-0.70mm}{\kern-0.70mm}{\kern-0.65mm}%
  {\kern-0.50mm}R}}

\font \bolditalics = cmmib10
\def\bx#1{\leavevmode\thinspace\hbox{\vrule\vtop{\vbox{\hrule\kern1pt
        \hbox{\vphantom{\tt/}\thinspace{\bf#1}\thinspace}}
      \kern1pt\hrule}\vrule}\thinspace}

\def \vc #1{{\textfont1=\bolditalics \hbox{$\bf#1$}}}
\def\eck#1{\left\lbrack #1 \right\rbrack}

\def\rund#1{\left( #1 \right)}

\def\D{{\cal D}}

\def\P{{\cal P}}

\def\d{{\rm d}}

\def\eps{{\epsilon}}

\def\vp{\varphi}

\def\Real{{\rm I\mathchoice{\kern-0.70mm}{\kern-0.70mm}{\kern-0.65mm}%
  {\kern-0.50mm}R}}
\def\C{\rm C\kern-.42em\vrule width.03em height.58em depth-.02em
       \kern.4em}
\font \bolditalics = cmmib10
\def\bx#1{\leavevmode\thinspace\hbox{\vrule\vtop{\vbox{\hrule\kern1pt
        \hbox{\vphantom{\tt/}\thinspace{\bf#1}\thinspace}}
      \kern1pt\hrule}\vrule}\thinspace}

\def \vc #1{{\textfont1=\bolditalics \hbox{$\bf#1$}}}
{\catcode`\@=11
\gdef\SchlangeUnter#1#2{\lower2pt\vbox{\baselineskip 0pt \lineskip0pt
  \ialign{$\m@th#1\hfil##\hfil$\crcr#2\crcr\sim\crcr}}}
}

\def\ueber#1#2{{\setbox0=\hbox{$#1$}%
  \setbox1=\hbox to\wd0{\hss$\scriptscriptstyle #2$\hss}%
  \offinterlineskip
  \vbox{\box1\kern0.4mm\box0}}{}}

\def\bx#1{\leavevmode\thinspace\hbox{\vrule\vtop{\vbox{\hrule\kern1pt
        \hbox{\vphantom{\tt/}\thinspace{\bf#1}\thinspace}}
      \kern1pt\hrule}\vrule}\thinspace}

\def\be{\begin{equation}}
\def\ee{\end{equation}}
\def\ba{\begin{eqnarray}}
\def\ea{\end{eqnarray}}

\def\d{{\rm d}}

\def\ltsima{$\; \buildrel < \over \sim \;$}
\def\lsim{\lower.5ex\hbox{\ltsima}}
\def\gtsima{$\; \buildrel > \over \sim \;$}
\def\gsim{\lower.5ex\hbox{\gtsima}}

\begin{document}



   \title[]{Cosmic variance of weak lensing surveys in the non-Gaussian regime}
\author[]
{\parbox[]{6.in}{Elisabetta Semboloni$^{1}$\thanks{sembolon@iap.fr}, Ludovic Van Waerbeke$^{2}$, Catherine Heymans$^{2}$, Takashi Hamana$^{3}$, Stephane Colombi$^{1}$, Martin White$^{4}$, Yannick Mellier$^{1,5}$\\
  \footnotesize $^1$ Institut d'Astrophysique de Paris, UMR7095 CNRS,
              Universit\'e Pierre~\&~Marie Curie - Paris, 98 bis bd Arago, 75014 Paris, France. \\
$^2$ University of British Columbia, 6224 Agricultural Road, Vancouver, V6T 1Z1, B.C.,
Canada\\
$^3$ National Astronomical Observatory of Japan Mitaka, Tokyo 181-8588, Japan\\
$^4$     Department of Astronomy, University of California, Berkeley, CA 94720-7300, United States.\\
   $^5$ Observatoire de Paris, LERMA, 61, avenue de
l'Observatoire,  75014 Paris, France.\\
}}
\maketitle
\begin{abstract}
The results from weak gravitational lensing analyses are  subject to a 
cosmic variance error term that has previously been  estimated assuming Gaussian 
statistics. In this letter we address the issue of estimating cosmic  
variance errors for weak lensing surveys in the non-Gaussian regime.\\Using 
standard cold dark matter
model ray-tracing simulations characterized by $\Omega_m=0.3, ~\Omega_\Lambda=0.7, ~h=0.7, ~\sigma_8=1.0$ for different  
survey redshifts $z_s$, we determine the variance of the two-point  shear correlation function measured across 64 independent lines of  sight.  We compare the measured variance to the variance expected  from a random Gaussian field and derive a redshift-dependent non-Gaussian calibration relation.\\We find that the ratio between the non-Gaussian and Gaussian  variance at 1 arcminute 
can be as high as $\sim 30$ for a survey with  source redshift $z_s \sim 0.5$ and $\sim 10$ for $z_s \sim 1$.  
The  transition scale $\vartheta_c$ above which the ratio is consistent  with unity, is found to be 
$\vartheta_c \sim 20$ arcmin for $z_s \sim  0.5$ and $\vartheta_c \sim 10$ arcmin for $z_s \sim 1$.  
We provide  fitting formula to our results permitting the estimation of non-Gaussian cosmic variance 
errors and  discuss the impact on current and future surveys.\\    
A more extensive set of simulations will however be required to  investigate the dependence of our results on cosmology, specifically on the amplitude of clustering . 
\end{abstract}
\begin{keywords}
cosmology: theory - gravitational lenses - large-scale structure 
\end{keywords}
\section{Introduction}
Weak lensing by large scale structure, i.e. cosmic shear, offers a direct
way of investigating the statistical properties of matter  in the
Universe, without making any assumptions on the relation between dark and
luminous matter.
 Current surveys are  large enough to provide high precision constraints
 on cosmology and the latest measurements performed with the Canada France Hawaii
 Telescope Legacy Survey
\cite{Hetal06,Setal06} is a step in that direction. Most of the
cosmological constraints from weak lensing use two-point shear
statistics \cite{R03,VWM03}, and  a crucial step in these
cosmological parameter measurements is the estimate of error bars
and systematics. Several papers address, statistically, the issue of
systematics from E and B modes \cite{CNPT01b,P02,SK06}, but only
few papers address the estimation of cosmic variance of cosmic shear
measurements \cite{WH99,CetH01,S02}. The latter assumes that the error
on the two-point shear correlation function follows Gaussian
statistics. However, we know that this is not the case at small scales
where non-linear effects become important. Cooray \& Hu 2001  use the dark matter halo model in Fourier space to study
non-Gaussian covariance. A tentative calibration of this effect on
the aperture mass statistic
 \cite{VW02}  showed that departure from
Gaussianity is expected to occur at  angular scales $\lsim 10$
arcminutes. The purpose of this {\it Letter} is to estimate the
non-linear covariance of the two-point shear correlation function in
real space, such that it can be of direct practical use for weak
lensing studies, as in Schneider et al. 2002, without having to calculate high order correlation functions  semi-analytically. Using
ray-tracing simulations for a model close to
the concordance cosmological model
\cite{Sp06} at different source redshift slices, we obtain a
redshift dependent calibration formula of the Gaussian covariance
derived in Schneider et al. 2002. This calibration takes the form of a matrix
with which the Gaussian covariance is multiplied by, to obtain the
non-Gaussian covariance.
This letter is organised as follows. The Section 2 provides the notation relevant for this work, and
the theoretical description of the Gaussian covariance.
Section 3 describes the ray-tracing simulations and
Section 4 shows our results. In Section 5 we show their impact on
current and future contiguous weak lensing surveys. We conclude by discussing the
limitation of our approach and the work that remains to be done in
order to achieve percent level accuracy in the non-linear covariance estimate.
\section{Cosmic Shear and Covariance}
We follow the notation of  Schneider et al. 1998. The power spectrum
$P_\kappa (k)$ of the convergence $\kappa$ is given by
\begin{eqnarray}
P_\kappa(k)&=&{9\over 4}\Omega_0^2\int_0^{w_H} {{\rm d}w \over
a^2(w)} P_{3D}\left({k\over f_K(w)};
w\right)\times\nonumber\\
&&\left[ \int_w^{w_H}{\rm d} w' n(w') {f_K(w'-w)\over
f_K(w')}\right]^2, \label{pofkappa}
\end{eqnarray}
where $f_K(w)$ is the comoving angular diameter distance out to a
distance $w$ ($w_H$ is the horizon distance), and $n(w(z))$ is the
redshift distribution of the sources. $P_{3D}(k)$ is the
3-dimension  non-linear mass power spectrum \cite{PD96,S03}, and
$k$ is the 2-dimension wave vector perpendicular to the
line-of-sight. We are interested in the non-Gaussian covariance
of the two-point shear correlation function, because it can be
easily transposed to other two-point statistics \cite{S02} by a suitable integration in  $k$-space.
The shear correlation function measured at angular scale $\vartheta$ can
be split into two components, $\xi_\pm $, where
\begin{equation}
\xi_{\pm}(\vartheta)={1\over 2\pi} \int_0^\infty {\rm d}
k~k~P_\kappa (k)~J_{0, 4} (k~\vartheta),
\end{equation}
and $J_{0, 4}$ is a Bessel function of the first kind, of zeroth order for
$\xi_+$ and of fourth order for $\xi_-$. The covariance matrix ${\rm Cov(\xi_+ ;\vartheta_1,\vartheta_2)}$ of the total shear correlation function
$\xi_+$ can be written as a sum of three different parts:
\begin{eqnarray}
{\rm Cov(\xi_+ ; \vartheta_1,\vartheta_2)}&=&\langle
\xi_+(\vartheta_1)\xi_+(\vartheta_2)\rangle\nonumber=\\D\delta_K(\vartheta_1-\vartheta_2)+q_{++}&+& <4^{th}\, \rm{order\, correlations }> \label{covdef}
\end{eqnarray}
 The first term is the diagonal statistical noise, depending on the intrinsic ellipticity variance, $\sigma_e$, the total area of the survey, $A$, and the density of galaxies, $n$. In practical units gives:
\begin{eqnarray}
D=3.979\times 10^{-9} \Big( \frac{\sigma_e}{0.3}\Big) ^4 \Big(\frac{A}{1 \, \rm{deg} ^2}\Big)^{-1}\times\nonumber\\\Big(\frac{n}{30 \, \rm{arcmin}^{-2}}\Big)^{-2}\Big(\frac{\theta}{1\, \rm{arcmin}}\Big) \Big(\frac{\Delta\theta/\theta}{0.1}\Big)^{-1}\label{noise}
\end{eqnarray}
where $\Delta\theta$ is the bin size used for the sampling of the correlation function. The second term represents the coupling between the noise and two point shear correlation function:
\be q_{++}={2\sigma_\eps^2\over \pi A n}
\int_0^\pi\d\vp\;\xi_+\rund{\sqrt{\vartheta_1^2+\vartheta_2^2-2\vartheta_1\vartheta_2\cos\vp}
}  \ee
and it can easily be calculated using a prediction for non-linear shear power spectrum \cite{PD96,S03}.
The third term  requires the knowledge of the fourth order shear correlation function as a function of scale.  If we assume Gaussian statistics, it can be expressed as a sum of two terms \cite{S02}:
\begin{eqnarray}
 r_{+0}&=&{2\over
\pi A}\int_0^\infty\d\phi\,\phi
\int_0^\pi\,\d\vp_1\,\xi_+(|\vc\psi_a|)
\int_0^\pi\,\d\vp_2\,\xi_+(|\vc\psi_b|) \;,
\nonumber \\
r_{+1}&=&{1\over (2\pi) A}\int_0^\infty\d\phi\,\phi
\int_0^{2\pi}\d\vp_1\,\xi_-(|\vc\psi_a|)\label{qrrdef} \\
&\times & \int_0^{2\pi}\d\vp_2\,\xi_-(|\vc\psi_b|) \eck{\cos 4\vp_a
\,\cos 4\vp_b + \sin 4\vp_a \,\sin 4\vp_b}\nonumber\;,
\end{eqnarray}
and $\vp_a$, $\vp_b$ are the polar angles of $\vc\psi_a$,
$\vc\psi_b$, respectively, $\cos
4\vp_a=1-8\psi_{a1}^2\psi_{a2}^2/|\vc\psi_a|^4$, $\sin
4\vp_a=4\psi_{a1}\psi_{a2}(\psi_{a1}^2 -\psi_{a2}^2)/|\vc\psi_a|^4$,
and the analogous expressions for $\vp_b$.

In this paper we are interested in the last term of eq.~(\ref{covdef}). At large scales we know that we can use the Gaussian approximation and write it as the sum of $r_{+0}$ and $ r_{+1}$. At small scales the Gaussian statistics break down and this term  cannot be calculated with semi-analytical techniques. The rest of the paper discusses our technique to calibrate the Gaussian prediction of this quantity in order to fit the non-Gaussian value measured in ray-tracing simulations. Therefore using ray-tracing simulations, we will measure the covariance of $\xi_+$, $\rm{Cov_{measured}(\xi_+;\vartheta_1,\vartheta_2)}$, assuming  $\sigma_e=0$, so $q_{++}=0$ and $D=0$  and we will define ${\cal F}(\vartheta_1,\vartheta_2)$, the ratio between the measured covariance matrix and Gaussian expectation for the  covariance matrix:
\begin{equation}
{\cal F}(\vartheta_1,\vartheta_2)={\rm
Cov_{measured}(\xi_+;\vartheta_1,\vartheta_2)\over
Cov_{Gaussian}(\xi_+;\vartheta_1,\vartheta_2)}  \label{covratio}
\end{equation}
where ${\rm{Cov_{Gaussian}}}(\xi_+;\vartheta_1,\vartheta_2)=r_{+0}+r_{+1}$.
\section{Description of the simulations}
\label{sec:simu}
We performed 16 particle in mesh (PM) dark matter simulations to cover a light cone of angular size $7\times7$ degrees,
from redshift $z=0$ to $z \simeq 3$,  using the tiling technique proposed by White \& Hu (2000)
  and explained in Appendix B of Hamana et al. (2002).
  We used 7 simulations of size 200 Mpc, 4 of size 400 Mpc, 3 of size 600 Mpc and 2 of size 800 Mpc.
Each $N$-body experiment involved $256^3$ particles in a grid of size $1024^3$ to compute
the forces.  The cosmology is  a standard $\Lambda$CDM model with $\Omega=0.3$, 
$\Omega_{\rm baryons}=0.04$, $\Lambda=0.7$ and $H_0=70$ km/s/Mpc, closed to the concordance
  model \cite{Sp06}, with a slightly higher value for the normalisation $\sigma_8=1$.
Combining the simulation data in different ways, we  generated 64 different, 
 albeit not fully independent (see below), light cones.
Each of them is divided in 64 successive redshift planes separated from each other
by 100 Mpc.
 The ray-tracing method is described in Hamana et al. (2002).
The spatial resolution of our simulations translates in an angular resolution of the
order of $\theta \simeq 0.5$ arcmin for $z \ga 0.2$. Given the limitations of the PM technique, 
 discreteness effects can be significant at redshift $z \ga 1.5$ (due to transcients). 
 Nevertheless, our measurements are reliable at scales larger than the mean interparticle distance, 
  i.e. $\theta \sim 2$ arcmin.  and we expect they can still used with high confidence level 
down to $\theta \simeq 1$ arcmin.

The size, $S$, of our light cones matches closely that of the simulations, so using
the dispersion among them to compute the covariance matrix would certainly
underestimate its amplitude, even at small angular scales.
  Fluctuations at scales larger than the simulation box size are also missed with
these realisations. Furthermore, they are not strictly independent, since they
just combine in different ways the 20 simulations. For these two reasons, in the case of $A=S$  the value of ${\cal F}(\vartheta_1,\vartheta_2)$  on small scales would be always underestimated, as compared to the cases $A \neq S$, and  would not converge to unity at large scales.
  In order to minimise these limitations and still have a fair estimate of the covariance matrix on 
  the estimator used here, it is thus wise to always keep 
the angular size of the survey $A$ to a small fraction of $S$. In practice,
we divide $S$ in 4, 9 and 16 adjacent subsamples, leading to assumed values
of $A \simeq 12$, $5.4$ and $3.1$ square degrees and 256, 576, 1024 realisations respectively, in total. 
Note that the choice of $A$ is made such that the largest
angular scale considered, $\theta=20$ arcmin, remains small compared to $\sqrt{A}$. We finally choose $A=5.44 ~{\rm deg^2}$.
\section{Description of the matrix calibration}
We measure ${\cal F}(\vartheta_1,\vartheta_2)$
according to eq.~(\ref{covratio}) as follows. The term ${\rm
Cov_{measured}(\xi_+;\vartheta_1,\vartheta_2)}$ is given
by $\langle (\xi_+ - \langle \xi_+ \rangle )^2 \rangle$,
where $\xi_+$ is measured in each realisation of the survey
of size $A=5.44~ {\rm deg^2}$, while the average $\langle \cdots \rangle$ is performed
over all the realisations. The term 
${\rm Cov_{Gaussian}(\xi_+;\vartheta_1,\vartheta_2)}$ is calculated
by measuring $\xi_+$ and $\xi_{-}$ in the 64 largest samples $S$
of area $A=49\ {\rm deg}^2$, and integrating numerically eqs.~(\ref{qrrdef}).
This ensures that the numerator and
denominator in eq.~(\ref{covratio}) are self-consistently defined.
It is worth noticing that  for all cases with $A<S $ the asymptotic behavior of  ${\mathcal F}(\vartheta_1,\vartheta_2)$ 
does not converge to unity. It indeed seems to be even worse than for the case $A =S$. This is a well-known effect which occurs when the scales become
comparable to the size of the survey (Peebles 1974). The result is that
for those scales the measured shear correlation is biased to lower values. 
 Therefore,  at small scales,   the  measured cosmic variance ${\rm Cov_{measured}(\xi_+;\vartheta_1,\vartheta_2)}$, when
 $A <S $ , is more  biased low and decreases faster when the scale increases  than for the case ${\rm Cov_{Gaussian}(\xi_+;\vartheta_1,\vartheta_2)}$ 
  and $ A = S $. 
The final result  is that the ratio ${\mathcal F}(\vartheta_1,\vartheta_2)$  becomes  smaller than unity. Note that in practice, for numerical reasons we have to use  
 $A = S $ to compute ${\rm Cov_{Gaussian}(\xi_+;\vartheta_1,\vartheta_2)}$   using \ref{qrrdef}. We do not expect this has 
any impact on our results, within the level of accuracy we can achieve from this set of simulations, provided we rescale the covariance matrix only in the inner part.
\begin{figure*}
\hbox{
  \psfig{figure=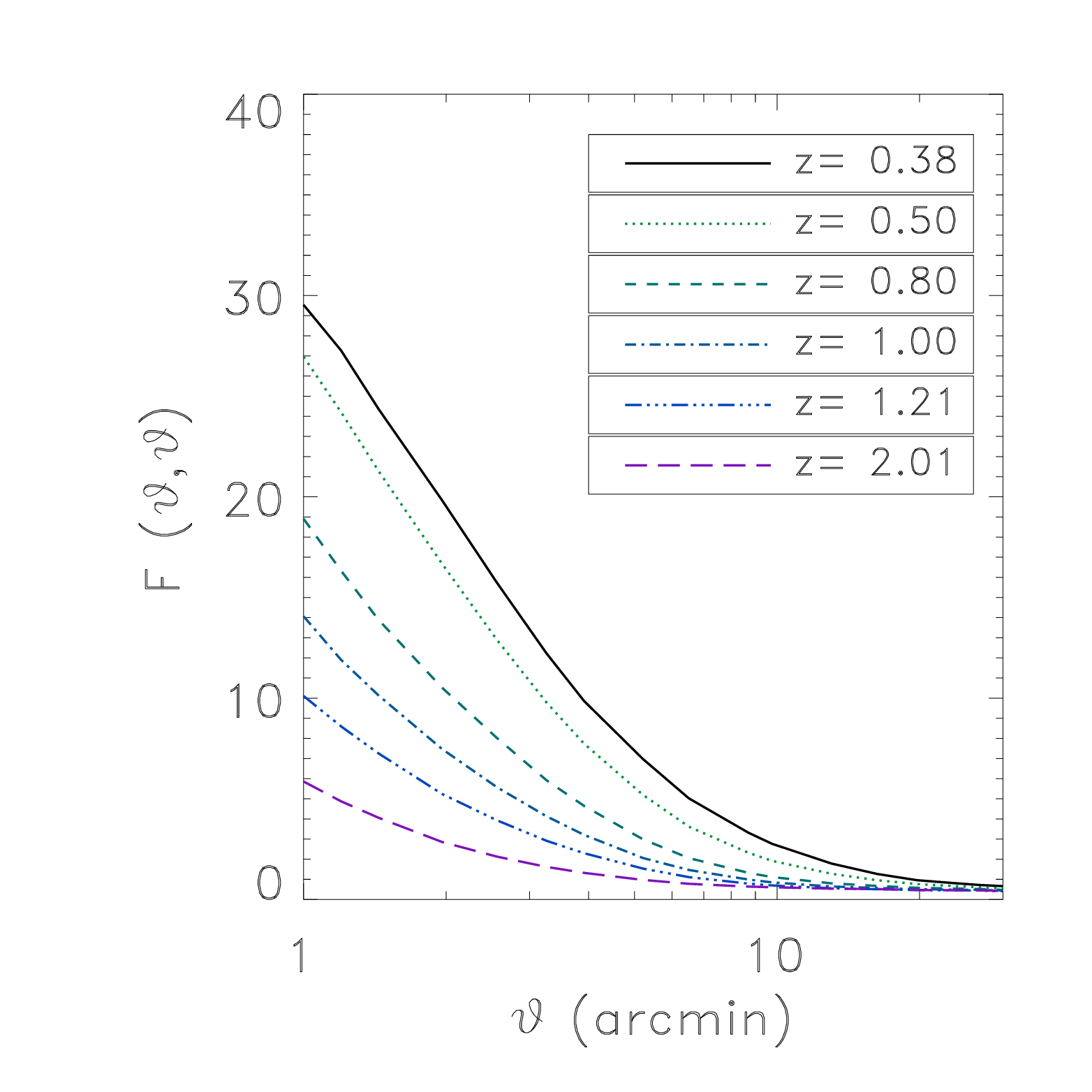,width=.45\textwidth,height=3.in}
  \psfig{figure=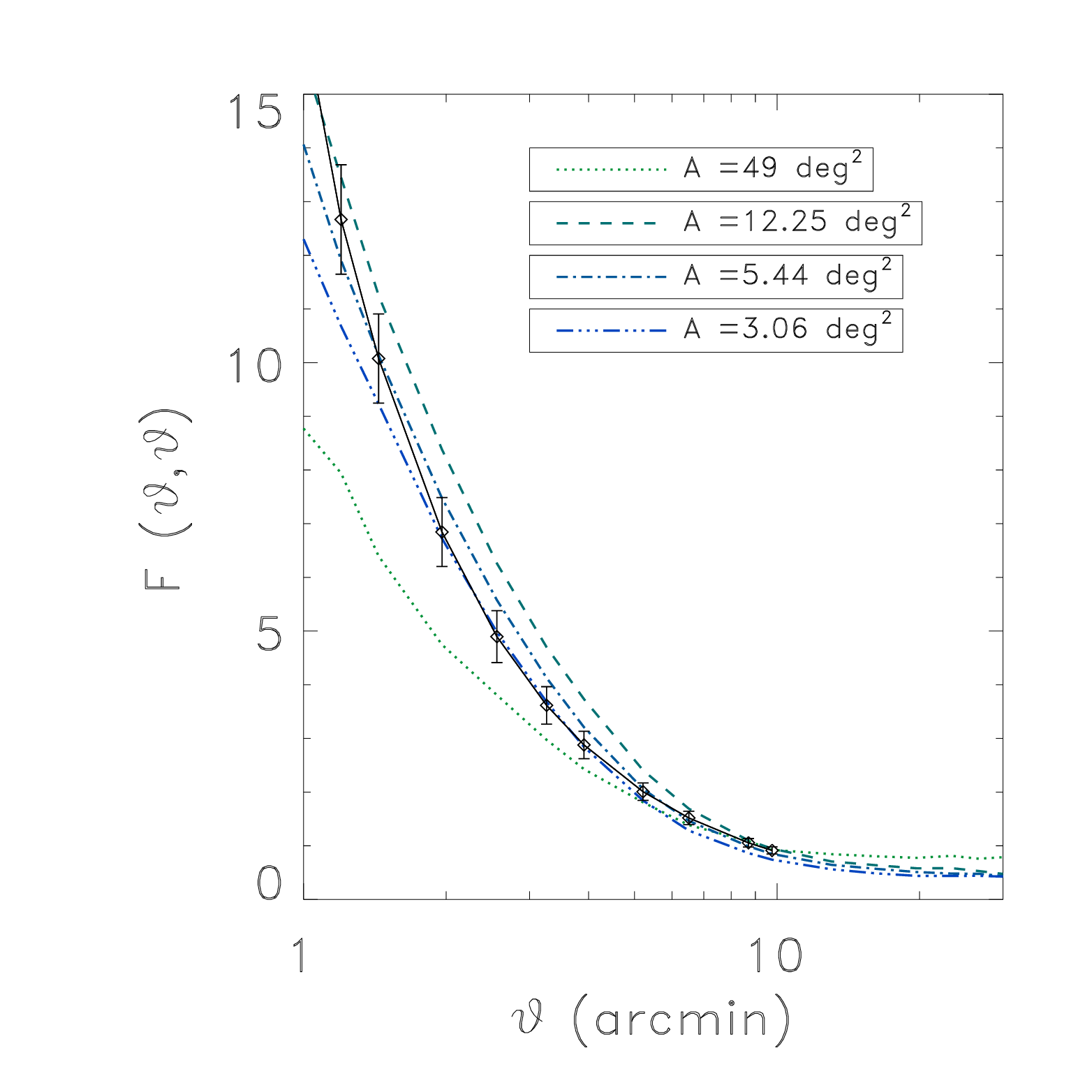,width=.45\textwidth,height=3.in}
}
\caption{Left panel: Diagonal elements of the matrix ${\cal F }(\vartheta_1,\vartheta_2)$ for different source redshift planes. Right panel:  Diagonal  elements ${\cal F}(\vartheta_1,\vartheta_2)$ for different survey sizes and  $z_s=1$. The black solid line represents the best-fit of ${\cal F }(\vartheta_1,\vartheta_2)$ using eq.~(\ref{param}). Error bars are computed using bootstrap with $1000$ realisations.}
\label{size}
\end{figure*}
\begin{figure*}
\hbox{
  \psfig{figure=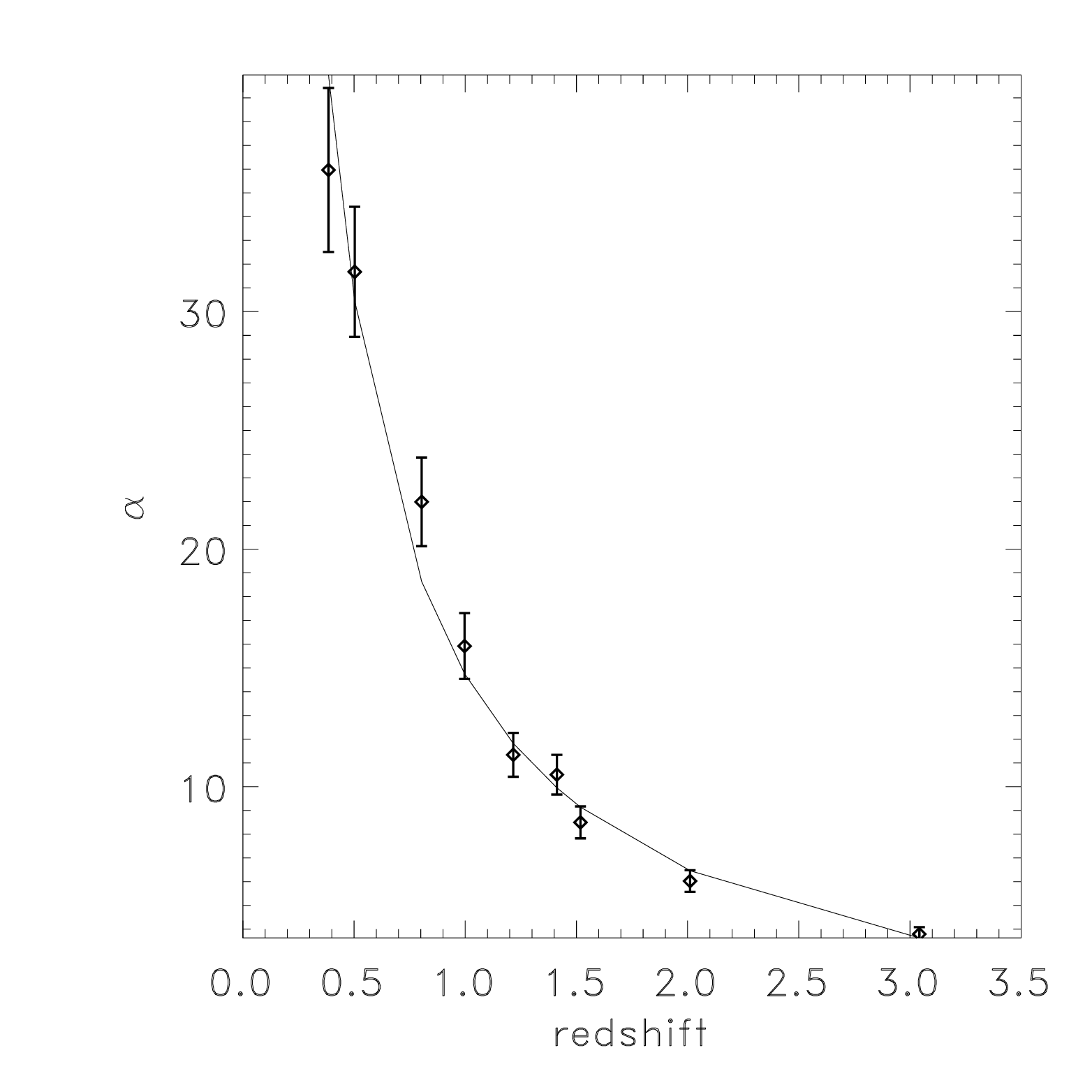,width=.33\textwidth}
  \psfig{figure=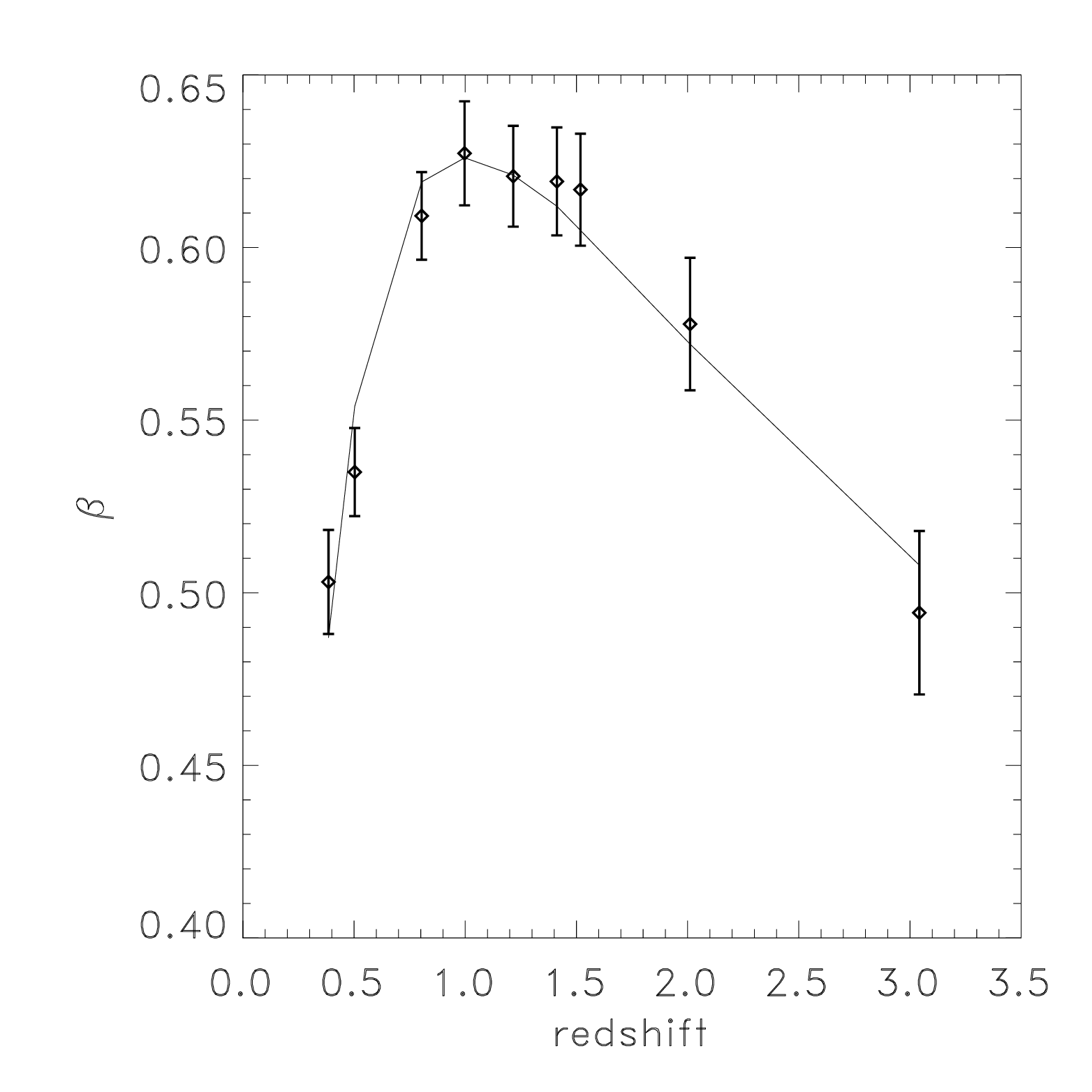,width=.33\textwidth}
  \psfig{figure=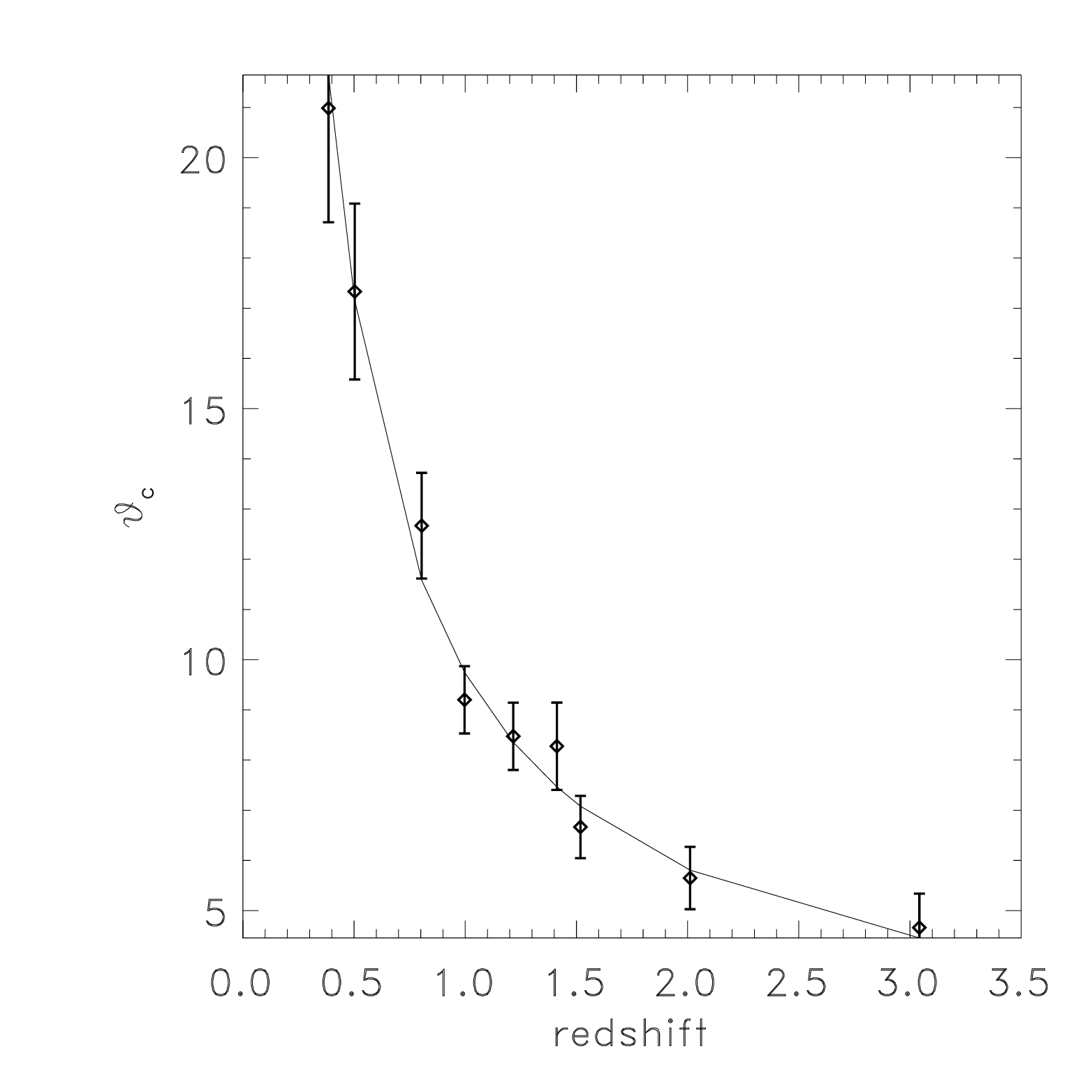,width=.33\textwidth} 
}
\caption{\label{p1p2_fit}
The plots shows  the measured parameters $\alpha$, $\beta$ and $\vartheta_c$ of the calibration matrix (see eq.~\ref{param}) as a function of the
redshift. Error bars are computed using bootstrap with 1000 realisations. The solid line shows the best fit from eq.~(\ref{pparam}). }
\end{figure*}
The left panel of Fig.\ref{size} shows the diagonal elements ${\cal
F}(\vartheta_1,\vartheta_2)$ for different source redshifts. For a
source redshift ${\rm z_{s}}\simeq 1$ the calibration factor is
$\sim 10$ at $\vartheta_1=\vartheta_2=1$ arcmin, 
implying that the cosmic variance has been widely
underestimated in previous lensing surveys at scales below $\sim 10$
arcminutes. The correction factor is larger for lower source
redshifts. The transition scale $\vartheta_c$, which defines the angular scale transition from
Gaussian and non-Gaussian covariance, is redshift dependent because the non-linear regime starts at larger scales for nearby structures. Therefore, the calibration
matrix must be parameterized with an explicit redshift dependence. We
choose a generic power law behavior, as suggested by the left panel of Fig.
\ref{size}, to parameterize ${\cal F}(\vartheta_1,\vartheta_2)$:

\begin{equation}
{\cal F}(\vartheta_1 ,\vartheta_2 )= {\alpha(z)\over \left[ \vartheta_1
\vartheta_2\right] ^{\beta(z)}}. \label{param}
\end{equation}
The two panels in Fig. \ref{p1p2_fit} show $\alpha$ and $\beta$ as
measured in the ray-tracing simulations at nine different source
redshifts $z_s=[0.4,0.5,0.8,1.0,1.2,1.4,1.5,2.0,3.0]$. These
measurements are well fit by the following redshift dependent
functions:
\begin{eqnarray}
\alpha(z)&=& {a_1\over z  ^{a_2}}+a_3 \nonumber\label{pparam}\\ \beta(z)&=& b_1~ z^
{b_2}~ {\rm exp}( -z ^ {b_3})+b_4. 
\end{eqnarray}
For $\alpha$, we find $(a_1,a_2,a_3)=(16.90,0.95,-2.19)$, and for
$\beta$, $(b_1,b_2,b_3,b_4)=(1.62,-0.68,-0.68,-0.03)$ in the
samples with angular size $A=5.4\ {\rm deg}^2$. 
 The fit is performed on scales below $10$ arcminutes, which allows
us to define also the transition angle $\vartheta_c$ as the scale
where the fitted function crosses the Gaussian covariance.  The third panel of 
Fig. \ref{p1p2_fit} represents the measurement of $\vartheta_c$. Using  the same 
functional form as for $\alpha$, namely $\vartheta_c={t_1\over z  ^{t_2}}+t_3$, 
We find the best fit values $(t_1,t_2,t_3)=(8.07,0.95,1.65)$.

 Since the normalisation of our simulations is high ($\sigma_8=1$),  we expect $\vartheta_c$ to be slightly overestimated.
 Several  other sources of uncertainty in our measurements
  may also spoil the estimate of the covariance.
In particular, as previously anticipated, there is  also a ``cosmic error'' and a ''cosmic bias'' that affect our measurements
(e.g. Szapudi \& Colombi 1996), which are
difficult to estimate.  Fortunately, such a cosmic bias/error is expected to increase
with the survey size $A$. According to eq.~(\ref{qrrdef}), the covariance scales as  $\propto 1/A$, so ${\cal F}$ should in fact be independent of $A$,
which allows one to use our parametrisation of ${\cal F}$ for any (reasonable)
angular survey size. This property can also be used to check the
convergence between our realisations of various survey sizes as illustrated
by right panel of Fig.~\ref{size}. Surveys with areas $A=3.1$, $5.4$ and $12$ square
degrees agree with each other, but there is a problem with $A=49\ {\rm deg}^2$,
where ${\cal F}$ is biased low. In the latter case, this is not surprising
since the light cone size is comparable to the simulations size, as discussed
in \S~\ref{sec:simu}. The convergence between other values of $A$ suggest that 
the cosmic bias/error on ${\cal F}$ measured in these samples is small, i.e.
the full set $S$ from which they are extracted, is a fair enough sample.
We check this by dividing our $A=12\ {\rm deg}^2$ set of 256 realisations into
4 subsamples of 64 realisations, and measured ${\cal F}$ in each of the subsamples. The dispersion
between these 4 subsamples is of the order of 10\% - 20\%, which gives a rough
idea of the accuracy of our estimate of ${\cal F}(\vartheta,\vartheta)$, in
agreement as well with the convergence between the measurements
observed on right panel of Fig.~\ref{size}
 for $A \leq 12\ {\rm deg}^2$. 

While our choice of parametrisation eq.~(\ref{param}) is globally accurate to $\sim
20\%$ along the diagonal of the matrix ${\cal
F}(\vartheta_1,\vartheta_2)$, it becomes less accurate for very
different $\vartheta_1$ and $\vartheta_2$. One should note that
the lack of accuracy in the off diagonal components  is not critical because the
cross-correlation coefficient is $\lsim 0.1$ in this region.


\section{Impact of non-Gaussianity on current and future Surveys}

\begin{figure}
\hbox{
\psfig{figure=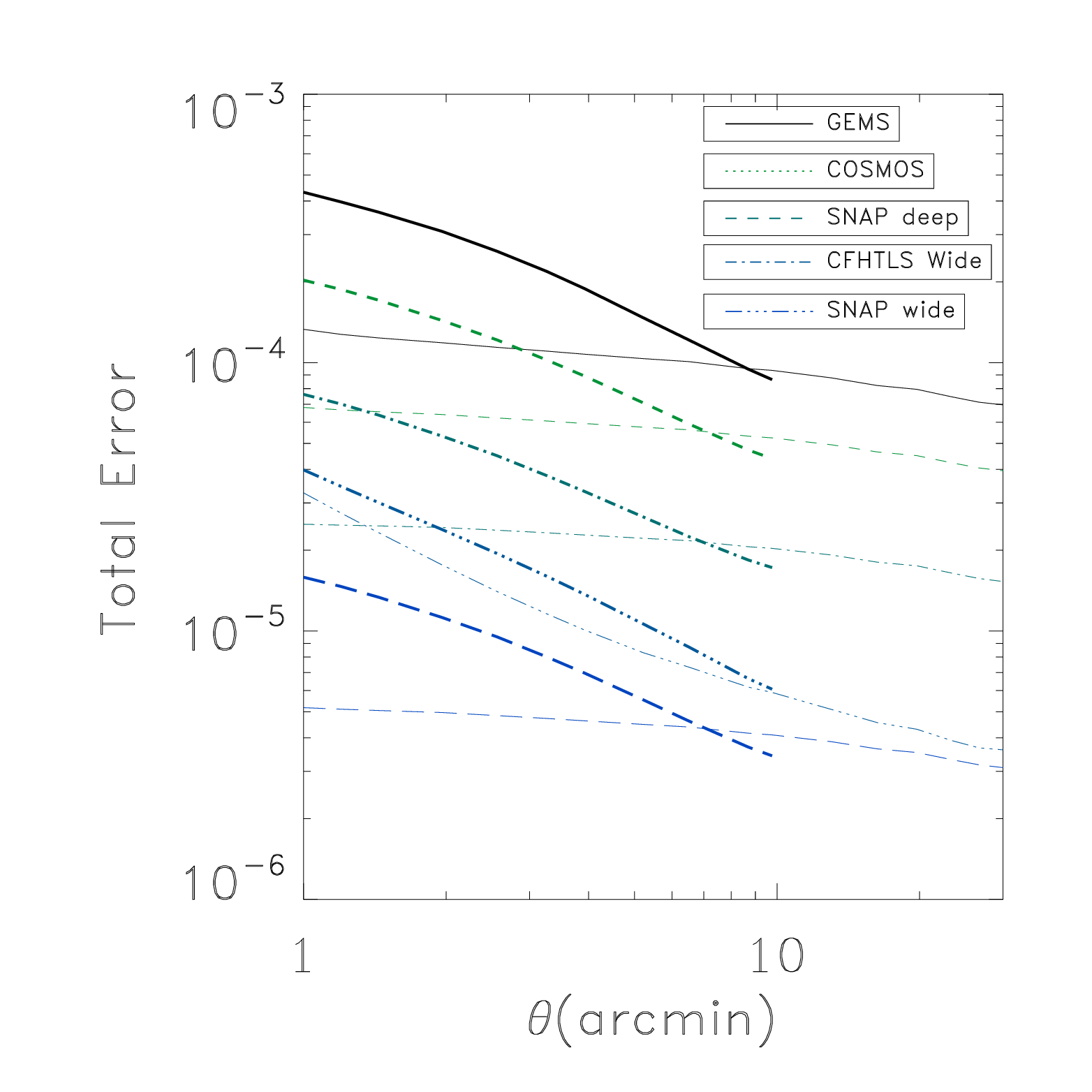,width=.45\textwidth,height=3.in}
}
\caption{\label{stat}
The total, statistical plus cosmic variance noise for each survey of table \ref{tab}. The noise including the non-Gaussian correction (thick lines) is compared on scales $ \vartheta \le 10 $  arcmin, with the noise expected in the case of Gaussian statistics (thin lines). Dropping the Gaussian assumption increases noise on small scales. The impact of the non-Gaussian  correction for the CFHTLS Wide is small; the statistical  noise $\propto 1/n^2$ and the coupling term $q_{++}\propto 1/n$ still dominates at small scales for such a density.  These same terms become negligible for the space based surveys whose density is much higher.  }
\end{figure}

\begin{figure}
\hbox{
\psfig{figure=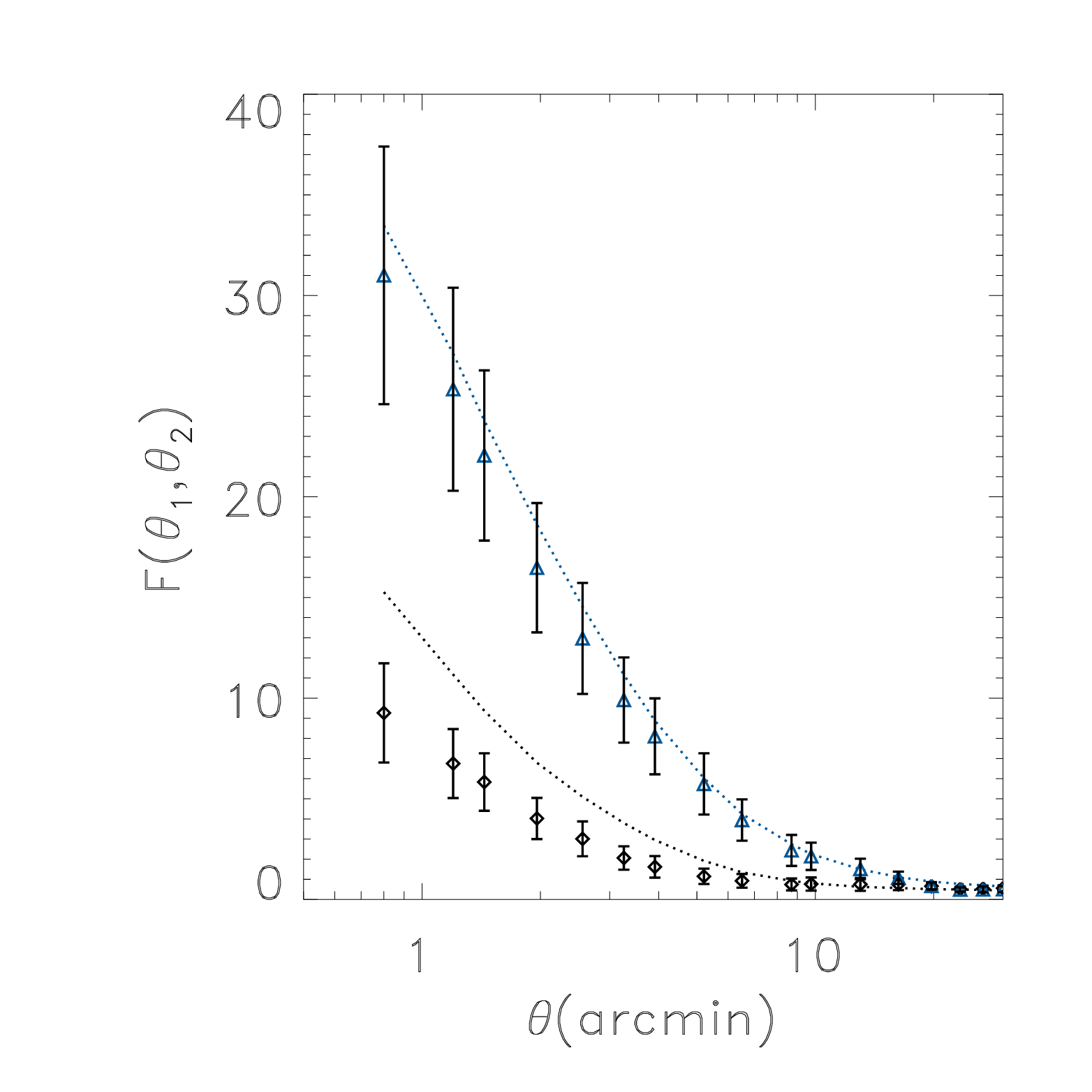,width=.45\textwidth,height=3.in}
}
\caption{\label{compare}  Diagonal elements of the matrix ${\cal F }(\vartheta_1,\vartheta_2)$ for planes with $z_s \simeq 0.5$ (blue triangles) and $z_s \simeq 1$ (black diamonds) obtained for the set of simulations $\Lambda CDM$ with $\sigma_8=0.8$. Error bars are obtained using bootstrap  with $1000$ realisations. For each of the two measurements  we compare the obtained value with the predicted value calculated from $\Lambda CDM$ simulations with $\sigma_8=1.$, marked with black and blue dotted lines.  }
\end{figure}

Finally, we compare the amplitude of statistical and cosmic variance  at small scales for a range of contiguous  surveys such as GEMS \cite{Hetal05}, COSMOS (Massey et al. in prep. ), CFHTLS Wide \cite{Hetal06} and two different versions of SNAP \cite{Retal04}  whose characteristics are shown in table \ref{tab}.
The statistical noise is computed using eq.~(\ref{noise}) , assuming a bin size  $\Delta \theta = 0.1$. Note that the statistical noise differs if the bin size used to measure the correlation function is different. In addition we choose $\sigma_e=0.4$ for ground-based surveys and $\sigma_e=0.3$ for space-based surveys.
Fig. \ref{stat} shows that by dropping the Gaussian approximation the total noise changes at small scales. The changing due to the non-Gaussian correction depends on the relative amplitude of the three different contributions to the total variance, namely, the shot noise, the sampling noise and the coupling term. For ``low density'' surveys, such as the CFHTLS Wide, the impact of the non-Gaussian correction is smaller as compared to the one expected  for the low noise space based surveys, where the cosmic variance far  exceeds the statistical noise. It is worth noticing our  results are obtained for a higher $\sigma_8$ value than Spergel et al. 2006 ($\sigma_8\simeq 0.75$) and are likely to be slightly different for this model.   A more extensive analysis of simulations made with different cosmologies would be  necessary to accurately  predict the amplitudes  of the non-Gaussianity corrections to the cosmic variance.
\begin{table}
\caption{Main Characteristics of surveys used in Fig. \ref{stat}.}\label{tab}
\begin{tabular}{|l|r|r|r|}
\hline
Name & A ($\rm{deg}^2$) & n & $<z_s>$\\
\hline
GEMS &  $0.25$ & $65$ &  $1$\\
COSMOS & $1.6$ & $80$ &  $1.2$\\
CFHTLS Wide & $50$ & $15$ & $0.8$  \\
$\rm{SNAP\,deep}$ & $15$ & $300$ & $1.4$ \\
$\rm{SNAP\,wide}$ & $260$ & $120$ & $1.2$ \\
\hline
\end{tabular}
\end{table}
Unfortunately, only a small set of  $\Lambda CDM$ ray tracing simulations  with $\sigma_8=0.8$ is available.
 This set of simulations, whose characteristics are given in Heymans et al. 2006, is  composed of  two redshift planes each containing  12 simulations of $25~{\rm deg^2}$ which is not enough to find a recalibration fitting formula. Nevertheless, Fig.\ref{compare}  shows that even for a  $\Lambda CDM$  model with  $\sigma_8=0.8$  the cosmic variance has been widely underestimated.
Fig.\ref{compare} also shows that using a rescaling obtained from $\sigma_8=1.0$ gives results which are in good agreement with the ones obtained for $\sigma_8=0.8$ for low redshift surveys and slightly overestimates the cosmic variance as the depth  increases. 
These simulations were also  used to confirm the validity of our statements  regarding the behavior of the ratio ${\mathcal F}(\vartheta_2,\vartheta_2)$  and  the change of the size of  $A$ used for the recalibration. 
\section{Discussion and Conclusion}
We have shown that the non-Gaussian contribution to the covariance
in  two-point shear statistics cannot be neglected at small
angular scales. Using ray-tracing simulations we have calibrated
the non-Gaussian covariance with respect to the Gaussian
covariance as calculated in Schneider et al. 2002. We have derived a
calibration matrix which can be used as a first approximation
for cosmological parameter measurements in current lensing surveys and for parameter
forecasting. 

We found that the correction coefficient could be as
high $10$ at $1$ arcminute for a source redshift of $1$, and $30$
for source redshift of $z_s=0.5$. The transition between Gaussian
and non-Gaussian covariance occurs around $10$ arcminutes for
$z_s=1$ and $20$ arcminutes for $z_s=0.5$.
Our work shows that it is important to include this
non-Gaussian contribution to the shear estimated errors, and that sub-arcminute
resolution ray-tracing simulations are  very useful for this
purpose. Although this source of error has been neglected in
previous lensing analysis, we note that it should not strongly
impact the measurement of $\sigma_8$ for surveys using the shear
signal measured above the transition scale $\vartheta_c$, where
the Gaussian covariance is a reasonable assumption. However, it
will significantly affect the joined $\Omega_M$-$\sigma_8$
constraints, since the degeneracy breaking between these two
parameters is based on a the relative amplitude of the shear correlation signal between small and
large scales \cite{JS97}. An increased error at small scale, as
shown here, will make the degeneracy more difficult to break. 

Extension of this work via a thorough analysis of the non-Gaussian
covariance based on numerical simulations include shear error
calibration with broad redshift distribution (tomography), different
two-points statistics  and the dependence of the non-Gaussian correction 
with a  varying cosmology. In particular we  expect a non-trivial
dependence of the calibration matrix with $\sigma_8$, since, for a
fixed angular scale, non-linear structures form earlier for higher
$\sigma_8$.
\section*{acknowledgements} We thank Peter Schneider for his constructive comments, 
ES thanks the hospitality of the University of British Columbia, which made this 
collaboration possible. LVW is supported by NSERC, CIAR CFI, CH is supported by a CITA national Fellowship,
 SC, YM and ES are supported by CNRS and PNC. 
This work was performed in part
within the Numerical Investigations in Cosmology group (NIC) as a task
of the HORIZON project.  The computational resources (NEC-SX5)  
for the present numerical simulations were made available
to us by the scientific council of the Institut de D\'eveloppement et
des Ressources en Informatique Scientifique (IDRIS).
This work has been supported in part by a Grant-in-Aid for
 Scientific Research (17740116) of the Ministry of Education,
 Culture, Sports, Science and Technology in Japan.
We thank the referee for his helpful comments.

\end{document}